\documentclass[doublespacing]{elsart}

\usepackage{graphicx}
\usepackage{amssymb}
\usepackage{epsf}

\newcommand\ltdash{\raise-1.8pt\hbox{$\scriptscriptstyle |$}}
\newcommand \beq  {\begin{equation}}
\newcommand \eeq  {\end{equation}}
\newcommand \bea {\begin{eqnarray} }
\newcommand \eea {\end{eqnarray}}

\newcommand\dg{^{\dagger}}
\newcommand\prl{\sl Phys. Rev. Lett. }
\newcommand\prb{ Phys. Rev. B}
\begin{document}
\begin{frontmatter}


\journal{SCES'2001: Version 1}

\title{
Pressure-Induced Magnetism and Hidden Order in $URu_2Si_2$
}
%
%
%
%
%
%

\author[NEC]{P. Chandra}
\author[RU] {P.Coleman}
\author[NL]{and J.A. Mydosh}
%
%

\address[NEC]{NEC Research, 4 Independence Way,
Princeton, NJ,USA
}
\address[RU]{Center for Materials Theory,
Department of Physics and Astronomy, 
Rutgers University, Piscataway, NJ 08854, USA.}
\address[NL]{
Kamerlingh Onnes Laboratory, 
Leiden University,
PO Box 9504, 2300 RA Leiden,
The Netherlands}

%
%
%
%


%
%
%
%

\begin{abstract}
We  discuss the  discovery of  pressure-induced  antiferromagnetism in
$URu_2Si_2$, in the context of  neutron, NMR and $\mu SR$ results. The
identification of a critical  pressure separating mean-field and Ising
phase transitions leads us to propose  that the system lies close to a
bicritical  point associated with  magnetic and  (non-magnetic) hidden
order.   We conclude  that  the recent  observation  of an  isotropic,
field-independent component in the silicon NMR line-width implies that
the hidden order parameter breaks time-reversal invariance and present
a preliminary discussion of the  underlying nature of the hidden order
parameter.
\end{abstract}

%
%



\end{frontmatter}

%
%
%
%
%
%
%
%
%
%
%
%
%
%
%
%
%
The origin of the entropy loss in $URu_2Si_2$  at $T_0 = 17.5 K$
is an outstanding
problem in heavy fermion physics.\cite{Buyers96} 
The observed 
staggered moment\cite{Broholm87} ($m_0 = 0.03 \mu_B$) cannot
account for the sharp discontinuities in bulk properties
\cite{Walter86,Maple86,Mason91,Miyako91,Ramirez92,Mason95} 
and the entropy change\cite{Palstra85}
that develop at this transition.
The field-dependences of
the magnetization and the gap\cite{Mentink96,vanDijk97} 
are dissimilar, implying the
presence of
a hidden order parameter $\psi$ that remains 
uncharacterized.\cite{Shah00,Miyako00}
 
Neutron experiments on $URu_2Si_2$ indicate that 
the staggered
magnetic moment increases linearly with applied
pressure,\cite{Amitsuka99}
$M\propto P$,
up to 1 GPa.
$M(T)$
is mean-field
for $P <$ 1 GPa
but is Ising in character at higher pressures.
In a parallel study,
NMR measurements\cite{Bernal01} on $URu_2Si_2$
show that at $T \le T_0$ the
silicon NMR line-width develops a field-independent, isotropic component
whose temperature-dependent magnitude is proportional to that of
the hidden
order parameter. These results imply 
an isotropic field distribution at the silicon
sites whose root-mean square value is
proportional to the hidden order
\begin{equation}\label{facts}
\langle B^{\alpha } (i) B^{\beta} (j)\rangle = a^{2}\psi ^{2} \delta
_{\alpha \beta }, 
\end{equation}
and is $\sim 10$ gauss at $T=0$.
This field magnitude is too
small to be explained by the observed moment which induces
a field
$B_{spin} = \frac{\mu_0 M}{a^2} = 100$ Gauss where
$a$ is the $U-U$ bond length ($a = 4 \times 10^{-10}$).
Furthermore this moment is aligned along the $c$-axis,
and thus cannot account for the isotropic nature
of the local field
distribution detected by NMR. 

Now we try to unify these two experiments
within a common framework. 
It has been widely assumed 
that the magnetic and the hidden
order coexist and are homogeneous.\cite{Shah00,Miyako00}
However recent NMR studies of $URu_2Si_2$ under
pressure\cite{Matsuda01}
indicate that for $T< T_0$ there is coexistence
of antiferromagnetic and paramagnetic regions,
implying that the magnetic and the hidden order
parameters are phase separated.\cite{Kivelson01}
Then the change in character of the magnetic transition
at $P=P_c$ is naturally
interpreted as originating from a bicritical point.  This picture, supported 
by $\mu SR$ data,\cite{Luke94} suggests that at ambient pressure
the observed magnetization is a volume fraction effect which
develops separately
from the hidden order via a first order transition.

In order to study this scenario further, we
assume that the free energy
$F[\psi,M,V]$ is function of 
the hidden order $\psi$, the staggered magnetization $M$
and the unit cell volume $V$ 
\begin{equation}
F = F_{
\psi }+F_{m}+
g \psi ^{2} M^{2}
\end{equation}
where
$F_{X}=(T_{X} (V)-T) X^{2 } + \frac{1}{2}
u_{X}^{2}X^{4}$ 
with $X = \{\psi,M\}$.
and $T_{\psi }=T_{M}$ at a critical volume $V_{c}$.  If 
$g^2 \ge u_{\psi}^{2} u_{M}^{2}$,
there exists a bicritical point\cite{Chaikin94} at $V = V_c$ 
with an associated first order
line as shown in Fig. 1a.  For $V > V_{c}$, the hidden order phase transition
temperature is stable and 
$T_{\psi }> T_{M}$.

In order to transform the $T-V$ phase diagram (Fig. 1a) into one for
$T-P$
(Fig. 1b),  we note that the 
pressure $P= -\frac{\partial F}{\partial V}$
is discontinuous across the 
first-order line in Fig. 1a leading to
two distinct pressure scales,
$P_{\psi}$ and $P_{M}$, in the $T-P$ plot (Fig. 1b). In the
associated
coexistence region (Fig. 1b),
the fraction of magnetic phase $x$
is given by the expression
$P (x) = (1-x) P_{\psi } + x P_{M}$
so that the  net magnetization is then 
\begin{equation}\label{mag}
{\mathcal  M}= Mx = M \left(\frac{P - P_{\psi } }{P_{M}- P_{\psi }} \right)
\end{equation}
Equation (\ref{mag}) displays the
linear development of $M(P)$ for $P > P_{\psi }$;
here we attribute a small $P_{\psi}$
to a large pressure-change associated with the first-order line in
Fig. 1a.  
In the hidden order phase, the magnetization
will develop via a first order phase transition 
in qualitative agreement with $\mu$SR data.\cite{Luke94}
Thus this simple approach
can model the observed change of the magnetization at low P. 

We now discuss the nature of the hidden order
parameter, emphasizing
the observed
isotropic field distribution at the silicon sites. 
The magnetic fields at these nuclei 
have two possible origins:\cite{Slichter78}
the electron-spin interaction 
and the orbital shift due to current
densities.
In $URu_{2}Si_{2}$, 
the electron fluid exhibits a strong Ising
anisotropy along the c-axis, as 
measured by the Knight shift\cite{Bernal01}; thus the electron-spin
interaction cannot be reponsible
for the isotropic fields at the silicon sites.  
Alternatively we propose that these 
local fields are induced by
currents that develop inside the crystal as the hidden order develops; thus
the observed isotropic line-width is attributed to the orbital shift.
We are therefore suggesting that for $T < T_0$, $URu_2Si_2$
is an orbital
antiferromagnet\cite{Halperin68}. Such states have been been
studied extensively in the context of the two-dimensional
Hubbard model,
\cite{Affleck88,Kotliar88,Nersesyan89,Schultz89},
particularly in connection with staggered flux
phases\cite{Wen96,Ivanov00}.  More recently
commensurate current density wave order 
has been proposed as an explanation 
of the spin-gap phase in the  underdoped
cuprate superconductors.\cite{Chakravarty00}

As a simple check on the applicability of
orbital antiferromagnetism to $URu_2Si_2$, we estimate 
local fields at the silicon
sites due to orbital currents circulating around
the square uranium plaquettes in the a-b plane.  On dimensional
grounds, the current along the $U-U$ bond is given by
$I = \frac{e \Delta }{\hbar }$
where $\Delta$  is the gap  associated with hidden  order formation. 
A 
microscopic derivation of this expression can be obtained from the Hubbard
model\cite{Hsu91}, assuming 
that the hidden order parameter \underline{is} the 
current along a $U-U$ bond. 
If this current loops around a  plaquette of side length
$\sim a$, then the field induced at a height $a$ above the
plaquette is approximately
$B = \left( \frac{\mu_{o}}{2 \pi a }\right )\left(\frac{e \Delta
}{\hbar } \right)$
Using the values $a= 4 \times 10^{-10}m$, $\Delta = 110K$, we obtain
$I=2.3 \mu A$ and 
$B= 11 \hbox{ Gauss}$, in good agreement with the local field strength
detected
in
$NMR$ and
$\mu SR$ measurements.\cite{Bernal01,Luke94}

In order to test whether this proposal
will yield local isotropic fields,
we allow the circulating
current around a plaquette (cf. Fig. 2)
centered at site $\bf  X$ to develop staggered order
${\mathcal I} ({\bf  X})~ = ~\psi e^{i {\bf Q}\cdot {\bf  X}}$.
The current along a bond is then the difference of the circulating
currents along its adjacent plaquettes.
The field at a silicon site can be computed using Ampere's law, where
the relevant vector potential is
\begin{equation}\label{vpotential}
{\bf A} ({\bf x})= \frac{\mu_{0} }{4\pi} \sum_j \int_{
{\bf x}_{j}^{(1)}
}^{
{\bf x}_{j}^{(2)}
}dx'\frac{{\bf I} ({\bf
x}_{j} )}{\vert {\bf x}- ( {\bf x}_{j}+{\bf  x}')\vert }.
\end{equation}
where ${\bf x}_{j}^{(1,2)}$ are the endpoints of the bond at site
${\bf x}_{j}$. 

The silicon atoms
in $URu_{2}Si_{2}$ are located at low-symmetry
sites, so that the fields do not cancel;
they reside above and
below the centers of the uranium plaquettes. 
The proposed orbital antiferromagnet 
must have ${\bf Q} \neq (\pi,\pi)$
in order to produce isotropic field distributions at the silicon
sites as displayed in Fig. 3a.
Using the vector potential in (\ref{vpotential}), our initial study finds that
an incommensurate $\bf  Q$ vector in the vicinity of $\bf Q=
(\frac{1}{4},\frac{1}{4},1)$ (Fig 3b) 
produces an isotropic field distribution at
the silicon sites. Such a configuration is staggered between the $U$
layers, with a periodicity of four unit cells in the basal plane. 

We end with a brief discussion about the microscopic nature of the
underlying hidden order.  First, the presence of isotropic local
fields\cite{Bernal01} at the silicon sites,
implies that $\psi$ must break time-reversal invariance.  Second,
we believe that the magnitude of the observed fields 
indicates that they are current-induced; this proposal is also
compatible with the observed robustness of $\psi$ to application of
high magnetic fields\cite{Mentink96}.
A simple possibility\cite{chandra01} is to identify the hidden order parameter
directly with a charge current,
corresponding to 
the imaginary part of an electron-hopping operator
\[
\vec{I}\propto -i \langle c\dg _{\sigma} ({\bf x}+\hat {\bf i}/2)
c_{\sigma} ({\bf x}-\hat {\bf i}/2)-\hbox{H.c.}\rangle .
\]
This bond current order would involve a significant fraction
of the entire
gap $\Delta = 110 K$ and thus could account for the observed local fields;
furthermore the associated entropy $\frac{S}{R \ln 2}~ =~
\frac{\Delta}{E_F}$ at $T_0$ could easily account for the experimentally
observed value of $\frac{S}{\ln 2} \sim 0.2 R$.  

Our proposal of incommensurate orbital antiferromagnetism in $URu_2Si_2$ can be
tested by experiment.  
$\mu SR$  measurements should confirm that the magnetic volume
fraction  of the sample increases with pressure. 
The incommensurate current ordering can be probed by
neutrons, whose scattering off the fields
produced by the orbital antiferromagnetism should produce (i) a small
incommensurate Bragg peak with 
a rapidly decaying form-factor characteristic of an extended object and (ii)
a dispersing gapless mode centered around the incommensurate Bragg
peak, associated with collective translations of the orbital antiferromagnet.

In summary, we have discussed the implications of three recent experiments
on $URu_{2}Si_{2}$.  We conclude that the observed pressure-induced antiferromagnetism
is probably due to phase separation.  We argue that the
development of isotropically distributed magnetic fields at the silicon
sites indicates that the hidden s
order parameter breaks
time-reversal invariance. Based on the size and isotropy 
of the measured local fields, we propose that $URu_2Si_2$ is
an incommensurate orbital antiferromagnet and make a number of predictions for experiment.  

We acknowledge discussions with 
G. Aeppli, O. Bernal, S. Chakravarty, G. Lonzarich and D. Morr.
Part of this work was performed at
the Aspen Center for Physics (P.C. and P.C) and
at Los Alamos National Laboratory (J.A.M).
This project is supported
under grant NSF-DMR 9983156 (Coleman).  

%
%
%
%

%
%
%
%


\vfill \eject 
\begin{figure}[here]
\begin{center}
\hbox{\epsfxsize=\textwidth \epsffile{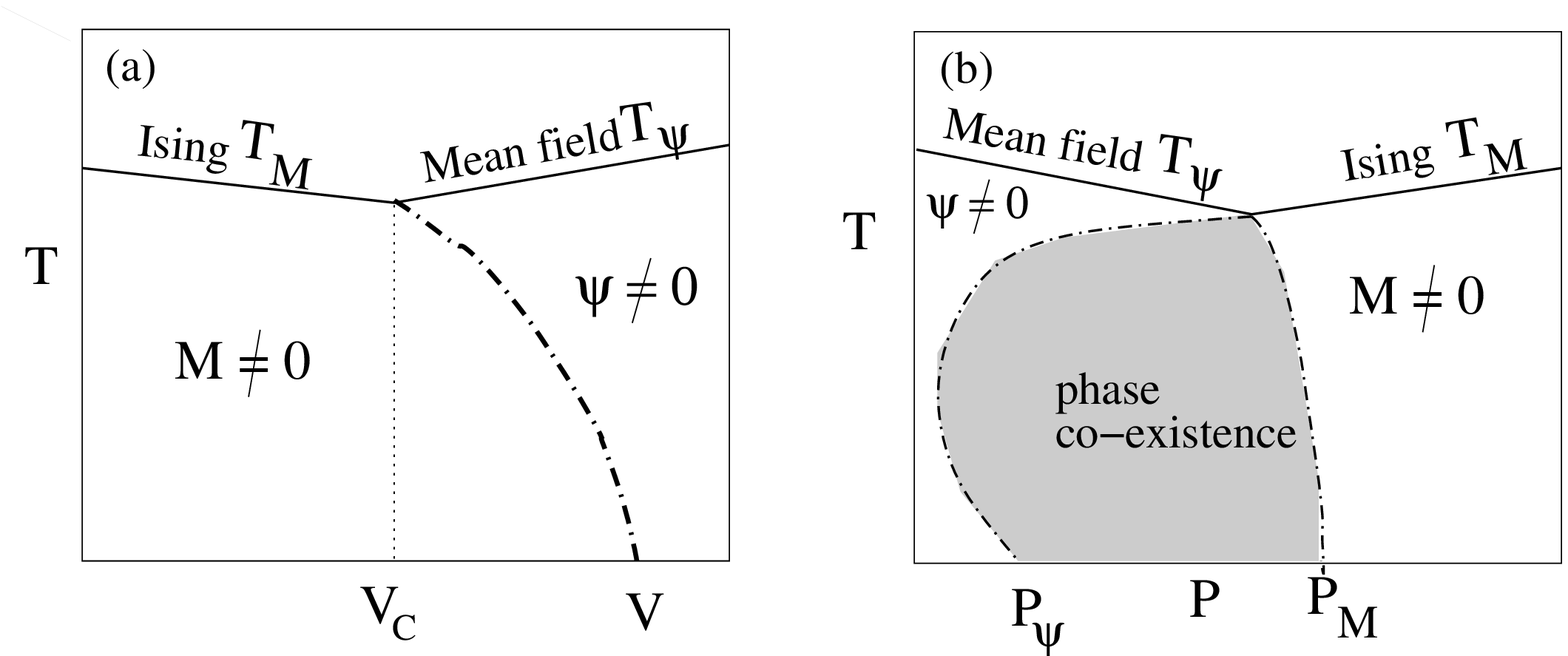}}
\end{center} 
    \caption{(a) Proposed temperature-volume phase diagram for
$URu_{2}Si_{2}$, with a first order line emanating from the
bicritical point where $T_{\psi }$ and $T_{M}$ are equal.
(b) In the temperature pressure diagram the first order line
is broadened into a region of phase co-existence, in which the
staggered moment is approximately linear in the applied pressure.
} 
 \end{figure}  
\vskip 0.4truein
 \begin{figure}[here]
\begin{center}
\hbox{\epsfxsize=0.8\textwidth \epsffile{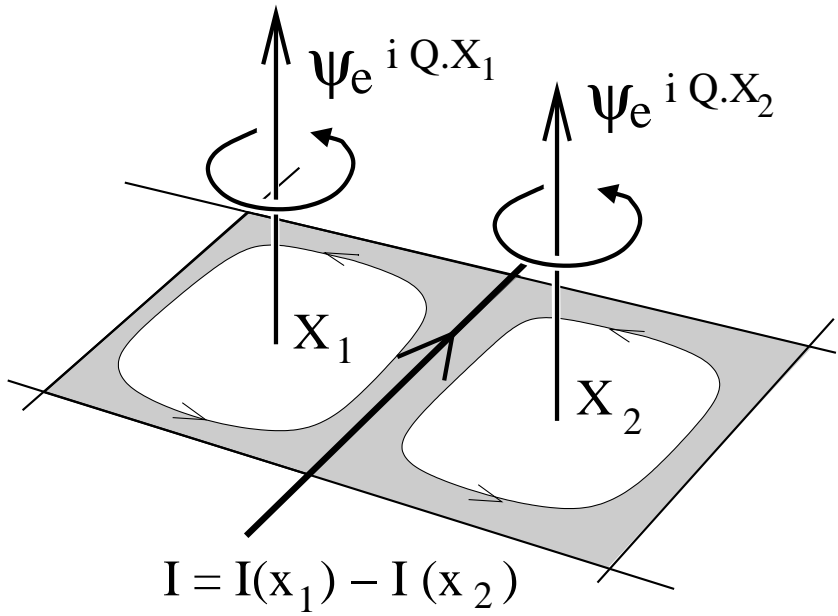}}
\end{center} 
     \caption{Schematic drawing showing how current along a bond
is the difference of the circulating currents around neighboring
Uranium plaquets.
} 
 \end{figure}  

 \begin{figure}[here]
\begin{center}
\hbox{\epsfxsize=0.8\textwidth \epsffile{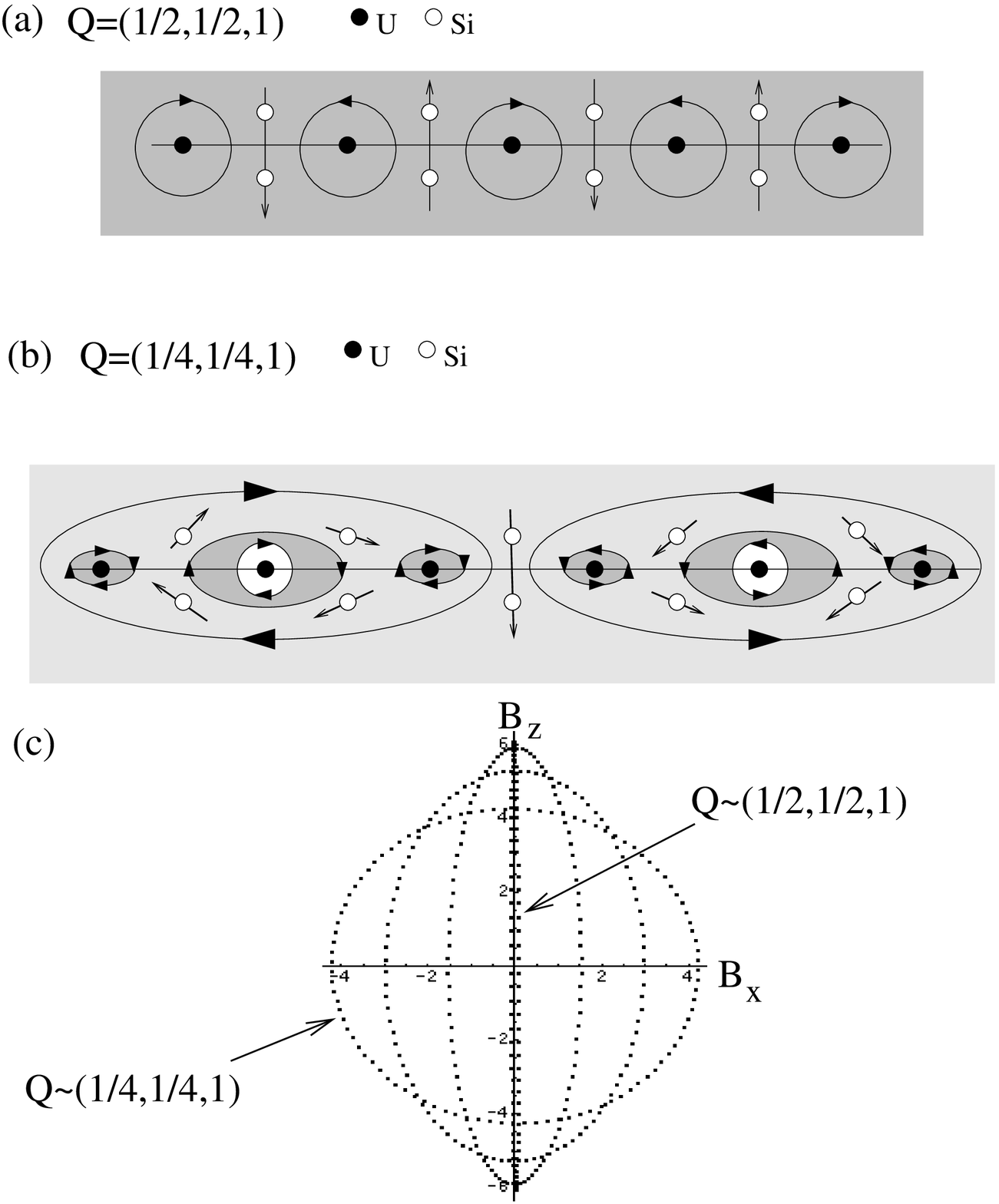}}
\end{center} 
     \caption{Schematic field distribution of (a) commensurate
and (b) incommensurate orbital antiferromagnet. Figure shows ``edge
on'' view along the a-axis of the unit cell.  (a) gives rise to an
anisotropic field distribution along the c-axes at the silicon sites. The 
$Q$ vector in (b) has been chosen to give an isotropic field
distribution at the silicon sites. (c) Showing field distributions
at Silicon sites for a sequence of four evenly spaced
$Q$ vectors between cases (a) and (b). 
} 
 \end{figure}

\end{document}